# A NEW APPROACH TO THE SOLUTION OF ECONOMIC DISPATCH USING PARTICLE SWARM OPTIMIZATION WITH SIMULATED ANNEALING

V.Karthikeyan[1] S.Senthilkumar[2] and V.J.Vijayalakshmi[3]

1Department of Electronics and communication engineering, SVSCE, Coimbatore India
`karthick77keyan@gmail.com`
2Department of Electronics and communication engineering, SVSCE, Coimbatore India
`sentheeyan@yahoo.co.in`
3Department of Electrical and electronics engineering, SKCET, Coimbatore, India
`vijik810@gmail.com`

### ABSTRACT

*Economic Dispatch is an important optimization task in power system. It is the process of allocating generation among the committed units such that the constraints imposed are satisfied and the energy requirements are minimized. More just, the soft computing method has received supplementary concentration and was used in a quantity of successful and sensible applications. Here, an attempt has been made to find out the minimum cost by using Particle Swarm Optimization (PSO) Algorithm using the data of three generating units. In this work, data has been taken such as the loss coefficients with the max-min power limit and cost function. PSO and Simulated Annealing (SA) are applied to find out the minimum cost for different power demand. When the results are compared with the traditional technique, PSO seems to give a better result with better convergence characteristic. All the methods are executed in MATLAB environment. The effectiveness and feasibility of the proposed method were demonstrated by three generating unit's case study. The experiment showed encouraging results, suggesting that the proposed approach of computation is capable of efficiently determining higher quality solutions addressing economic dispatch problems.*

## 1. INTRODUCTION

Power systems should be operated under a high degree of economy for competition of deregulation. Unit commitment is an important optimization task addressing this crucial concern for power system operations. Since Economic Dispatch (ED) is the fundamental issue during unit commitment process, it should be important to obtain a higher quality solution from ED efficiently. In essence, the objective of ED is to minimize total generation costs, while satisfying power demands and constraints [1] Previous solutions to ED problems have applied various mathematical programming methods and optimization techniques, such as the Lambda-iteration method [2], base point and participation factors method [3], the gradient method [4], and the Newton method [2]. Unfortunately, these methods contained essential assumptions which were the incremental costs of the generators should be monotonically increasing functions [2]. These assumptions proved to be infeasible for practical implementation because of their nonlinear characteristics of prohibited zones or valve-point effects in real generators. Therefore, dynamic programming [5], nonlinear programming [6], and their modification techniques for solving ED issues have been presented. However, these methods may cause dimensional problem when applied to modern power systems due to the large number of generators. In the past decade, several computational algorithm techniques, such as Genetic Algorithms (GAs) [7], [8], Simulated Annealing (SA) [9], Tabu Searches (TS) [10], and Artificial Neural Networks (ANNs) [11], [12] encompass been second-hand to tackle power optimization subjects. These algorithms are in the form of probabilistic heuristics, with global search properties. Though GA methods have been employed successfully to solve complex optimization problems, recent research has identified deficiencies in its performance. This degradation in efficiency is apparent in applications with highly epistatic objective functions (i.e., when optimized parameters are highly correlated), thereby, hampering crossover and mutation operations and compromising the improved fitness of offspring because population chromosomes contain similar structures. In addition, their average fitness becomes high toward the end of the evolutionary process [17]. Moreover, the premature convergence of GA degrades its performance by reducing its search capability, leading to a higher probability of being trapped to a local optimum [22]. Recently, a global optimization technique called Particle Swarm Optimization (PSO) [13]–[15] has been used to solve real time issues and aroused researchers' interest due to its flexibility and efficiency. Limitations of the classic greedy search technique, which restricts allowed forms of fitness functions, as well as continuity of the variables used, can be entirely eliminated. The PSO, first introduced by Kennedy and Eberhart [16], is a modern heuristic algorithm developed through the simulation of a simplified social system. It was found to be robust in solving continuous nonlinear optimization problems [19], [21]. In general, the PSO method is faster than the SA method because the PSO contains parallel search techniques. However, similar to the GA, the main adversity of the PSO is premature convergence, which might occur when the particle and group best solutions are trapped into local minimums during the search process. Localization occurs because particles have the tendency to fly to local, or near local, optimums, therefore, particles will concentrate to a small region and the global exploration ability will be weakened. On the contrary, the most significant characteristic of SA is its probabilistic jumping property, called the metropolis process. However, by adjusting the temperature, the metropolis process can be controlled. It has been theoretically proven that the SA technique converges asymptotically to the global optimum solution with probability, ONE [18], [19], provided that certain conditions are satisfied. Therefore, a novel SA-PSO approach is proposed in this paper. The salient features of PSO and SA are hybrid to create an innovative approach, which can generate high-quality solutions within shorter calculation times and offers more stable convergence characteristics. Moreover, to consider the nonlinear characteristics of a generator, such as prohibited operating zones and valve-point effects for actual power system operations, an effective coding scheme for particle representation is also proposed to prevent constraints violation during the SA-PSO process. The feasibility of the proposed method was demonstrated on four different systems and then compared with the real-coded PSO [15], GA [8], [30], and evolutionary algorithm [29] methods regarding solution quality and computational efficiency.

## 2. PROBLEM FORMULATIONS

The objective of ED is to minimize the total generation costs of a power system over an appropriate period (usually one hour), while satisfying various constraints. Cost efficiency is the most important sub problem of power system operations. Due to the highly nonlinearity characteristics of power systems and generators, ED belongs to a class of nonlinear programming optimization methods containing equality and inequality constraints. Practically speaking, while the scheduled combined units for each specific period of operation are listed from unit commitment, the ED planning must perform the optimal generation dispatch among the operating units in order to satisfy the system load demands and practical operation constraints of generators, which include ramp rate limits, maximum and minimum limits, and prohibited operating zones. Generally, the generation cost function is usually expressed as a quadratic polynomial. However, it is more practical to consider the valve-point effects for fossil-fuel-based plants. Therefore, the cost function considered in this paper can be represented as follows: The Let Ci mean the cost, expressed for example in dollars per hour, of producing energy in the generator unit I. The total controllable system production cost therefore will be

$$C = \sum_{i=1}^{N} C(i) \tag{1}$$

The generated real power $P_{Gi}$ accounts for the major influence on $C_i$. The individual real generation are raised by increasing the prime mover torques, and this requires a cost of increased expenditure of fuel. The reactive generations $Q_{Gi}$ do not have any measurable influence on $c_i$ because they are controlled by controlling by field current. The individual production cost $c_i$ of generators unit I is therefore for all practical purposes a function only of $P_{Gi}$, and for the overall controllable production cost, we thus have

$$C = \sum_{i=1}^{N} C(P_{Gi}) \tag{2}$$

When the cost function C can be written as a sum of terms where each term depends only upon one independent variable.

## 2.1. System Constraints

Broadly speaking there are two types of constraints
i) Equality constraints
ii) Inequality constraints

The inequality constraints are of two types (i) Hard type and, (ii) Soft type. The hard type are those which are definite and specific like the tapping range of an on-load tap changing transformer whereas soft type are those which have some flexibility associated with them like the nodal voltages and phase angles between the nodal voltages, etc. Soft inequality constraints have been very efficiently handled by penalty function methods.

### 2.1.1 Equality Constraints

From observation we can conclude that cost function is not affected by the reactive power demand. So the full attention is given to the real power balance in the system. Power balance requires that the controlled generation variables $P_{Gi}$ abbey the constraints equation shown in (3)

$$P_d = \sum_{i=1}^{N} C_i(P_{Gi}) \tag{3}$$

### 2.1.2 Inequality Constraints

The Inequality Constraints of various cases are given below.

### 2.1.2.1 Generator Constraints

The KVA loading in a generator is given by $\sqrt{P^2 + Q^2}$ and this should not exceed a pre specified value of power because of the temperature rise conditions

- The maximum active power generation of a source is limited again by thermal consideration and also minimum power generation is limited by the flame instability of a boiler. If the power output of a generator for optimum operation of the system is less than a pre-specified value P min , the unit is not put on the bus bar because it is not possible to generat that low value of power from the unit. Hence the generator power P cannot be outside the range stated by the inequalit $P_{min} \leq P \leq P_{max}$
- Similarly the maximum and minimum reactive power generation of a source is limited. The maximum reactive power is limited because of overheating of rotor and minimum is limited because of the stability limit of machine. Hence the generator powers Pp cannot be outside the range stated by inequality, i.e. $Q_{p\ min} \leq Q_p \leq Q_{p\ max}$

### 2.1.2.2 Voltage Constraints

It is essential that the voltage magnitudes and phase angles at various nodes should vary within certain limits. The normal operating angle of transmission lies between 30 to 45 degrees for transient stability reasons. A lower limit of delta assures proper utilization of transmission capacity.

### 2.1.2.3 Running Spare Capacity Constraints

These constraints are required to meet
a) The forced outages of one or more alternators on the system and
b) The unexpected load on the system

The total generation should be such that in addition to meeting load demand and losses a minimum spare capacity should be available i.e. G    $P_p + P_{so}$
Where G is the total generation and PSO is some pre-specified power. A well planned system is one in which this spare capacity PSO is minimum.

### 2.1.2.4 Transmission Line Constraints

The flow of active and reactive power through the transmission line circuit is limited by the thermal capability of the circuit and is expressed as,

$$C_p \quad C_p\ max$$

Where Cp max is the maximum loading capacity of the line.

### 2.1.2.5 Network security constraints

If initially a system is operating satisfactorily and there is an outage, may be scheduled or forced one, It is natural that is an outage, may be scheduled or forced one, it is natural that some of the constraints of the system will be violated. The complexity of these constraints (in terms of 10 numbers of constraints) is increased when a large system is under study. In this a study is to be made with outage of one branch at a time and then more than one branch at a time. The natures of constraints are same as voltage and transmission line constraints.

## 3. PROPOSED METHOD

### 3.1 PSO - An Optimization Tool

Particle swarm optimization (PSO) is a population based stochastic optimization technique developed by Dr.Ebehart and Dr. Kennedy in 1995, inspired by social behaviour of bird flocking or fish schooling. PSO shares many similarities with evolutionary computation techniques such as Genetic Algorithms (GA). The system is initialized with a population of random solutions and searches for optima by updating generations. However, unlike GA, PSO has no evolution operators such as crossover and mutation. In PSO, the potential solutions, called particles, fly through the problem space by following the current optimum particles. The detailed information will be given in following sections. Compared to GA, the advantages of PSO are that PSO is easy to implement and there are few parameters to adjust.

### 3.2 Background of Simulated Annealing

For certain problems, Simulated Annealing (SA) may be more efficient than exhaustive enumeration — provided that the goal is merely to find an acceptably good solution in a fixed amount of time, rather than the best possible solution. The name and inspiration come from annealing in metallurgy, a technique involving heating and controlled cooling of a material to increase the size of its crystals and reduce their defects. The heat causes the atoms to become unstuck from their initial positions (a local minimum of the internal energy) and wander randomly through states of higher energy; the slow cooling gives them more chances of finding configurations with lower internal energy than the initial one. By analogy with this physical process, each step of the SA algorithm replaces the current solution by a random "nearby" solution, chosen with a probability that depends both on the difference between the corresponding function values and also on a global parameter T (called the temperature), that is gradually decreased during the process. The dependency is such that the current solution changes almost randomly when T is large, but increasingly "downhill" (for a minimization problem) as T goes to zero. The allowance for "uphill" moves potentially saves the method from becoming stuck at local optima—which are the bane of greedier methods. The method was independently described by Scott Kirkpatrick, C. Daniel Gelatt and Mario P. Vecchi in 1983 and by Vlado erný in 1985. The method is an adaptation of the Metropolis-Hastings algorithm, a Monte Carlo method to generate sample states of a thermodynamic system, invented by M.N. Rosenbluth in a paper by N. Metropolis et al. in 1953.

### 3.3 Particle Swarm Optimization

PSO simulates the behaviours of bird flocking. Suppose the following scenario: a group of birds are randomly searching food in an area. There is only one piece of food in the area being searched. All the birds do not know where the food is. But they know how far the food is in each iteration. So what's the best strategy to find the food? The effective one is to follow the bird, which is nearest to the food. PSO learned from the scenario and used it to solve the optimization problems. In PSO, each single solution is a "bird" in the search space. We call it "particle". All of particles have fitness values, which are evaluated by the fitness function to be optimized, and have velocities, which direct the flying of the particles. The particles fly through the problem space by following the current optimum particles. PSO is initialized with a group of random particles (solutions) and then searches for optima by updating generations. In every iteration, each particle is updated by following two "best" values. The first one is the best solution (fitness) it has achieved so far. (The fitness value is also stored.) This value is called pbest. Another "best" value that is tracked by the particle swarm optimizer is the best value, obtained so far by any particle in the population. This best value is a global best and called g-best. When a particle takes part of the population as its topological neighbours, the best value is a local best and is called p-best. After finding the two best values, the particle updates its velocity and positions with following equation (4) and (5).

$$V_i^{(u+1)} = w * V_i^{(u)} + C_1 * rand()* (pbest_i - P_i^{(u)}) + C_2 * rand() * (gbest_i - P_i^{(u)}) \qquad (4)$$

$$P_i^{(u+1)} = P_i^{(u)} + V_i^{(u+1)} \qquad (5)$$

Where The term rand ( )* ($pbest_i - P_i^{(u)}$) is called particle memory influence.
The term rand( ) * ($gbest_i - P_i^{(u)}$) is called swarm influence.

$V_i^{(u)}$ is the velocity of $i^{th}$ particle at iteration 'u' must lie in the range $V_{min}$ $V_i$ $V_{max}$
In general, the inertia weight *w* is set according to the following equation,

$$W = Wmax - \left[\frac{Wmax - Wmin}{ITERmax}\right] \times ITER \qquad (6)$$

Where w -is the inertia weighting factor
$W_{max}$ - maximum value of weighting factor
$W_{min}$ - minimum value of weighting factor
ITERmax - maximum number of iterations
ITER - current number of iteration

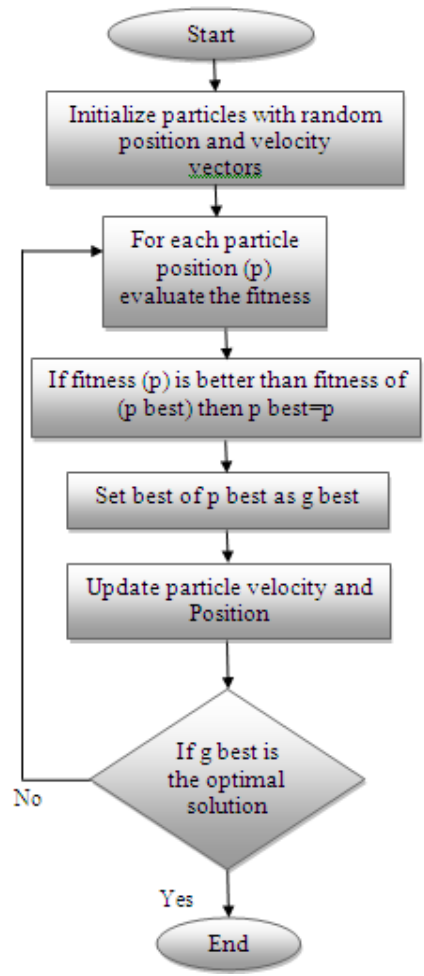

Figure1. Flow Chart for Particle Swarm Optimization

## 4. SIMULATED ANNEALING AND PSO

Simulated annealing (SA) is a generic probabilistic metaheuristic for the global optimization problem of locating a good approximation to the global optimum of a given function in a large search space. It is often used when the search space is discrete (e.g., all tours that visit a given set of cities). For certain problems, simulated annealing may be more efficient than exhaustive enumeration — provided that the goal is merely to find an acceptably good solution in a fixed amount of time, rather than the best possible solution. Recently there have been significant research efforts to apply evolutionary computation (EC) techniques for the purposes of evolving one or more aspects of Simulated Annealing. Evolutionary computation methodologies have been applied to three main attributes of Simulated Annealing.

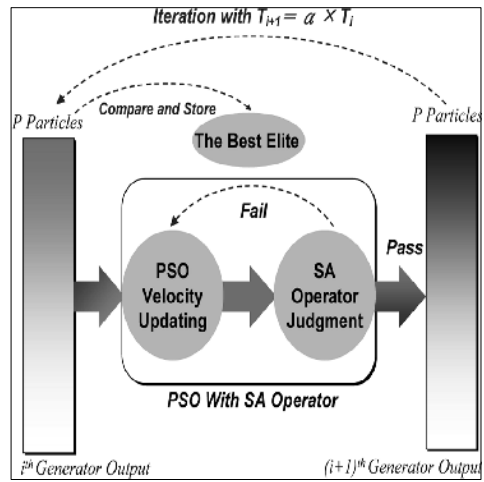

Figure 2. Concept Diagram of the Proposed Algorithm

Most of the work involving the evolution of SA has focused on the network weights and topological structure. The selection of fitness function depends on the research goals. For a classification problem, the rate of misclassified patterns can be viewed as the fitness value. The advantage of the EC is that EC can be used in cases with non-differentiable PE transfer functions and no gradient information available. The disadvantages are

1. The performance is not competitive in some problems.
2. Representation of the weights is difficult and the genetic operators have to be carefully selected or developed.
It is faster and gets better results in most cases.

## 5. IMPLEMENTATION OF THE PROPOSED METHOD

Particle Swarm optimization (PSO) is a population based algorithm in which each particle is considered as s solution in the multimodal optimization space. There are several types of PSO proposed but here in this work very simplest form of PSO is taken to solve the Economic Dispatch (ED) problem. The particles are generated keeping the constraints in mind for each generating unit. When economic load dispatch problem considered it can be classified in two different ways.

1. Economic load dispatch without considering the transmission line losses
2. Economic load dispatch considering the transmission line losses.

### 5.1 Economic Dispatch without loss

When any optimization process is applied to the ED problem some constraints are considered. In this work two different constraints are considered. Among them the equality constraint is summation of all the generating power must be equal to the load demand and the inequality constraint is the powers generated must be within the limit of maximum and minimum active power of each unit. The sequential steps of the proposed PSO method are given below.

Step 1:
The individuals of the population are randomly initialized according to the limit of each unit including individual dimensions. The velocities of the different particles are also randomly generated keeping the velocity within the maximum and minimum value of the velocities. These initial individuals must be feasible candidate solutions that satisfy the practical operation constraints.
Step 2:
Each set of solution in the space should satisfy the equality constraints .So equality constraints are checked. If any combination doesn't satisfy the constraints then they are set according to the power balance equation.
Step 3:
The evaluation function of each individual Pgi, is calculated in the population using the evaluation function F .Here F is

$$F = a \times (P_{gi})^2 + b \times P_{gi} + c \qquad (7)$$

Where a, b, c are constants. The present value is set as the pbest value.
Step 4:
Each pbest values are compared with the other pbest values in the population. The best evaluation value among the p-bests is denoted as gbest.
Step 5:
The member velocity v of each individual Pg is modified according to the velocity update equation

$$V_{id}^{(u+1)} = w * V_i^{(u)} + C_1 * rand(\ ) * (pbest_{id} - P_{gid}^{(u)}) + C_2 * rand(\ ) * (gbest_{id} - P_{gid}^{(u)}) \qquad (8)$$

Where u is the number of iteration
Step 6:
The velocity components constraint occurring in the limits from the following conditions are checked

$$Vd^{min} = -0.5 * P_{min}$$
$$Vd^{max} = +0.5 * P_{max}$$

Step 7:
The position of each individual $P_g$ is modified according to the position update equation

$$P_{gid}^{(u+1)} = P_{gid}^{(u)} + V_{id}^{(u+1)} \qquad (9)$$

Step 8:
If the evaluation value of each individual is better than previous *pbest,* the current value is set to be *pbest*. If the best *pbest* is better than *gbest*, the value is set to be *gbest*.
Step 9:
If the number of iterations reaches the maximum, then go to step 10.Otherwise, go to step 2.
Step 10:
    The individual that generates the latest gbest is the optimal generation power of each unit with the minimum total generation cost.

### 5.2 Economic Dispatch with loss

When the losses are considered the optimization process becomes little bit complicated. Since the losses are dependent on the power generated of the each unit, in each generation the loss changes, The P-loss can be found out by using the equation

$$P_L = \sum_m \sum_n P_m B_{mn} P_n \qquad (10)$$

Where $B_{mn}$ the loss co-efficient and the loss co-efficient can be calculated from the load flow equations or it may be given in the problem. However in this work for simplicity the loss coefficient are given which are the approximate one. Some parts are neglected. The sequential steps to find the optimum solution are

Step 1:
The power of each unit, velocity of particles, is randomly generated which must be in the maximum and minimum limit. These initial individuals must be feasible candidate solutions that satisfy the practical operation constraints.
Step 2:
Each set of solution in the space should satisfy the following equation

$$\sum_{i=1}^{N} P_{gi} = P_D + P_L \qquad (11)$$

$P_L$ calculated by using above equation (6.4). Then equality constraints are checked. If any combination doesn't satisfy the constraints then they are set according to the power balance equation.

$$P_d = P_D + P_L - \sum_{\substack{i=1 \\ i \neq d}}^{N} P \qquad (12)$$

Step 3:
The cost function of each individual $P_{gi}$, is calculated in the population using the evaluation function F. Here F is

$$F = a \times (P_{gi})^2 + b \times P_{gi} + c \qquad (13)$$

Where a, b, c are constants. The present value is set as the pbest value.
Step 4:
Each pbest values are compared with the other pbest values in the population. The best evaluation value among the pbest is denoted as *gbest*.
Step 5:
The member velocity v of each individual Pg is updated according to the velocity update equation.

$$V_{id}^{(u+1)} = w * V_i^{(u)} + C_1 * rand(\ ) * (pbest_{id} - P_{gid}^{(u)}) + C_2 * rand(\ ) * (gbest_{id} - P_{gid}^{(u)}) \qquad (14)$$

Where u is the number of iteration
Step 6:
The velocity components constraint occurring in the limits from the following conditions are checked

$$V_{d\,min} = -0.5 * P_{min}$$
$$V_{dmax} = +0.5 * P_{max}$$

Step 7:
The position of each individual Pg is modified according to the position update equation

$$P_{gid}^{(u+1)} = P_{gid}^{(u)} + V_{id}^{(u+1)} \qquad (15)$$

Step 8:
    The cost function of each new is calculated. If the evaluation value of each individual is better than previous *pbest*; the current value is set to be *pbest*. If the best *pbest* is better than *gbest*, the value is set to be *gbest*.

Step 9:
    If the number of iterations reaches the maximum, then go to step 10.Otherwise, go to step 2.

Step 10:
    The individual that generates the latest gbest is the optimal generation power of each unit with the minimum total generation cost.

## 6. RESULTS AND DISCUSSION

The proposed coding scheme for the solution of Economic Dispatch using Particle Swarm Optimization algorithm and Simulated Annealing technique has been tested on the three unit system. Here we have taken C1 = 1.99 and $C_2$ = 1.99 and the maximum value of w is chosen 0.9 and minimum value is chosen 0.4.the velocity limits are selected as $V_{max}$= 0.5*$P_{max}$ and the minimum velocity is selected as $V_{min}$= -0.5*$P_{min}$. There are 10 number of particles selected in the population. The cost curve (the plot between the number of iterations and the cost) for the three unit system considered here is given below. For different value of $C_1$ and $C_2$ the cost curve converges in the different region. So, the best value is taken for the minimum cost of the problem. If the numbers of particles are increased then cost curve converges faster. It can be observed the loss has no effect on the cost characteristic. The simulation output is shown in Fig.3.

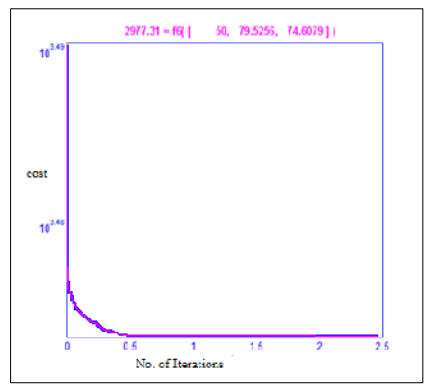

Figure 3. Simulation Result - Cost Curve

## 7. CONCLUSIONS

This paper presents a new approach to address practical ED issues. Power crisis is one of the major issues of concern all over the world today. The production is not enough to meet the demands of consumers. Under these circumstances the power system should be efficient in Economic Load Dispatch which minimizes the total generating cost. This project presents a new approach to address practical ED issues. A new approach to the solution of ED using Particle Swarm Optimization with Simulated Annealing has been proposed, and proven by a systematic simulation processes. Through the proposed coding scheme, constraints of ED problems can be effectively released during the search process, therefore, the solution quality, as well as the calculation time, is greatly improved. The proposed approach has been demonstrated by three unit system and proven to have superior features, including high quality solutions, stable convergence characteristics, and good computational efficiency. The generation limits and the demand are considered for practical use in the proposed method. The encouraging simulation results showed that the proposed method is capable of obtaining more efficient, higher quality solutions for ED problems. In future, the most problematic line flow constraints and security constraints will be considered which leads to the Security Constrained Economic Dispatch (SCED). For handling the line flow constraints, in addition to the Particle Swarm Optimization (PSO) algorithm with Simulated Annealing(SA), the Newton Raphson method (NR) will be used as well as for handling the security constraints also. Furthermore, the generating cost will be minimized and the Security Constrained Economic Dispatch will be targeted.

## REFERENCES

[1] Cheng-Chien Kuo, "A Novel Coding Scheme for Practical Economic Dispatch by Modified Particle Swarm Approavh," *IEEE Trans. Power Syst.*, vol. 23, no. 4, Nov. 1990.
[2] B. H. Chowdhury and S. Rahman, "A review of recent advances in economic dispatch," *IEEE Trans. Power Syst.*, vol. 5, no. 4, pp. 1248–1259, Nov. 1990.
[3] A. J. Wood and B. F. Wollenberg, *Power Generation, Operation and Control* New York: Wiley, 1984.
[4] C. L. Chen and C. L. Wang, "Branch-and-bound scheduling for thermal generating units," *IEEE Trans. Energy Convers.*, vol. 8, no. 2, pp. 184–189, Jun. 1993.
[5] K. Y. Lee *et al.*, "Fuel cost minimize at ion for both real- and reactive power dispatches," *Proc. Inst. Elect. Eng., Gen., Tranm. Distrib.* vol. 131, no. 3, pp. 85–93, 1984
[6] Z. X. Liang and J. D. Glover, "A zoom feature for a programming solution to economic dispatch including transmission losses," *IEEE Trans. Power Syst.*, vol. 7, no. 3, pp. 544–550, Aug. 1992.
[7] A. M. Sasson, "Nonlinear programming solutions for load-flow, minimum- loss, and economic dispatching problems," *IEEE Trans. Power App. Syst.*, vol. PAS-88, no. 4, pp. 399–409, Apr. 1969.
[8] D. C. Walters and G. B. Sheble, "Genetic algorithm solution of economic dispatch with valve point loading," *IEEE Trans. Power Syst.*, vol. 8, no. 3, pp. 1325–1331, Aug. 1993.
[9] P. H. Chen and H. C. Chang, "Large-Scale economic dispatch by genetic algorithm," *IEEE Trans. Power Syst.*, vol. 10, no. 4, pp. 1919–1926, Nov. 1995.
[10] K. P. Wong and C. C. Fung, "Simulated annealing based economic dispatch algorithm," *Proc. Inst. Elect. Eng., Gen., Tranm. Distrib.* vol. 140, no. 6, pp. 509–515, Nov. 1993.
[11] W. M. Lin, F. S. Cheng, and M. T. Tsay, "An improved tabu search for economic dispatch with multiple minima," *IEEE Trans. Power Syst.*, vol. 17, no. 1, pp. 108–112, Feb. 2002.
[12] R. H. Liang, "A neural-based re-dispatch approach to dynamic generation allocation," *IEEE Trans. Power Syst.*, vol. 14, no. 4, pp. 1388–1393, Nov. 1999.
[13] T. Yalcinoz and M. J. Short, "Neural networks approach for solving economic dispatch problem with transmission capacity constraints," *IEEE Trans. Power Syst.*, vol. 13, no. 2, pp. 307–313, May 1998.
[14] A. I. Selvakumar and K. Thanushkodi, "A new particle swarm optimization solution to nonconvex economic dispatch problems," *IEEE Trans. Power Syst.*, vol. 22, no. 1, pp. 42–51, Feb. 2007.
[15] J. B. Park, K. S. Lee, J. R. Shin, and K. Y. Lee, "A particle swarm optimization for economic dispatch with nonsmooth cost functions," *IEEE Trans. Power Syst.*, vol. 20, no. 1, pp. 34–42, Feb. 2005.
[16] Z. L. Gaing, "Particle swarm optimization to solving the economic dispatch considering the generator constraints," *IEEE Trans. Power Syst.*, vol. 18, no. 3, pp. 1187–1195, Aug. 2003.
[17] J. Kennedy and R. Eberhart, "Particle swarm optimization," in *Proc. IEEE Int. Conf. Neural Networks*, 1995, vol. 4, pp. 1942–1948.
[18] R. C. Eberhart and Y. Shi, "Comparison between genetic algorithms and particle swarm optimization," in *IEEE Int. Conf. Evolutionary Compututation*, May 1998, pp. 611–616.
[19] D. Mitra, F. Romeo, and A. Sangiovanni-Vincentelli, "Convergence and finite-time behavior of simulated annealing," in *Proc. 24th Conf. Decision and Control*, Dec. 1985, pp. 761–767.
[20] M. Locatelli, "Convergence properties of simulated annealing for continuous global optimization," *J. Appl. Probab.*, vol. 33, pp. 1127–1140, 1996.
[21] Y. Shi and R. Eberhart, "A modified particle swarm optimizer," in *IEEE Int. Conf. Evolutionary Computation*, May 1998, pp. 69–73
[22] Y. Shi and R. C. Eberhart, "Empirical study of particle swarm optimization," in *Proc. Congr. Evolutionary Computation*, 1999, pp. 1945–1950
[23] D. B. Fogel, *Evolutionary Computation: Toward a New Philosophy of Machine Intelligence*, 2 ed. Piscataway, NJ: IEEE Press, 2000.
[24] Z. L. Gaing, "Particle swarm optimization to solving the economic dispatch considering the generator constraints," *IEEE Trans. Power Syst.*, vol. 16, no. 3, pp. 1187–1195, Aug. 2003.

[25] S. Kirkpatrick, C. D. Gelatt, and M. P. Vecchi, "Optimization by simulated annealing," *Science*, vol. 220, pp. 671–680, 1983
[26] N. Metropolis, A. W. Rosenbluth, M. N. Rosenbluth, and A. H. Teller, "Equation of state calculations by fast computer machines," *J. Chem. Phys.*, vol. 21, no. 6, pp. 1087–1092, 1953.
[27] T. Yalcionoz, H. Altun, and M. Uzam, "Economic dispatch solution using a genetic algorithm based on arithmetic crossover," in *Proc. IEEE Porto Power Tech. Conf.*, Porto, Portugal, Sep. 2001.
[28] H. Yoshida, K. Kawata, Y. Fukuyama, S. Takayama, and Y. Nakanishi, "A particle swarm optimization for reactive power and voltage control considering voltage security assessment," *IEEE Trans. Power Syst.*, vol. 15, no. 4, pp. 1232–1239, Nov. 2000.
[29] F. N. Lee and A. M. Breipohl, "Reserve constrained economic dispatch with prohibited operating zones," *IEEE Trans. Power Syst.*, vol. 8, no. 1, pp. 246–254, Feb. 1993.
[30] J. O. Kim, D. J. Shin, J. N. Park, and C. Singh, "Atavistic genetic algorithm for economic dispatch with valve point effect," *Elect. Powe r Syst. Res.*, vol. 62, no. 3, pp. 201–207, Jul. 2002.
[31] A. Pereira-Neto, C. Unsihuay, and O. R. Saavedra, "Efficient evolutionary strategy optimization procedure to solve the nonconvex economic dispatch problem with generator constraints, Generation," *Proc. Inst. Elect. Eng., Gen., Tranm., Distrib.*, vol. 152, no. 5, pp. 653–660, Sep. 2005.
[32] W. Y. Ng, "Generalized generation distribution factors for power system security evaluations," *IEEE Trans. Power App. Syst.*, vol. PAS-100, pp. 1001–1005, Mar. 1981.